%% file: skeleton.tex
\documentclass[a4paper,11pt]{article}
\usepackage{pos}
\usepackage{outlines}
\usepackage{graphicx}

\usepackage{lineno}
\usepackage{placeins}
\usepackage{subcaption}
\usepackage{comment}

\title{Search for high-energy neutrinos from magnetars with IceCube}
 \ShortTitle{High-energy neutrinos from magnetars}

\author{The IceCube Collaboration \\{\normalsize \normalfont(a complete list of authors can be found at the end of the proceedings)}\\}

\emailAdd{aghadimi@crimson.ua.edu}
\emailAdd{jmsantander@ua.edu}

\abstract{Neutron stars with very strong magnetic fields are known as magnetars. There are multiple theories that predict magnetars may be able to emit high-energy (HE) neutrinos through hadronic processes by accelerating cosmic rays to high energies. A subclass of magnetars known as soft gamma-ray repeaters (SGRs) can produce giant flares that can result in the production of HE neutrinos. Some magnetars also exhibit bursting activity during which they may emit HE neutrinos. Here we describe our time-integrated search for neutrino emission from magnetars listed in the McGill Online Magnetar Catalog and three newly discovered magnetars SGR 1830-0645, Swift J1555.5-5402, and NGC 253. SGR 1830-0645 and Swift J1555.2-5402 were discovered in 2020 and 2021 respectively by SWIFT after emitting short bursts. A very bright short gamma-ray burst that is believed to be a magnetar giant flare has been localized to NGC 253. We use 14 years of well-reconstructed muon-neutrino candidate events collected by the IceCube Neutrino Observatory to look for significant clustering in the direction of magnetars.

\vspace{4mm}
{\bfseries Corresponding authors:}
Ava Ghadimi$^{1*}$, Marcos Santander$^{1}$\\
{$^{1}$ \itshape University of Alabama}\\[4mm]
$^*$ Presenter

\ConferenceLogo{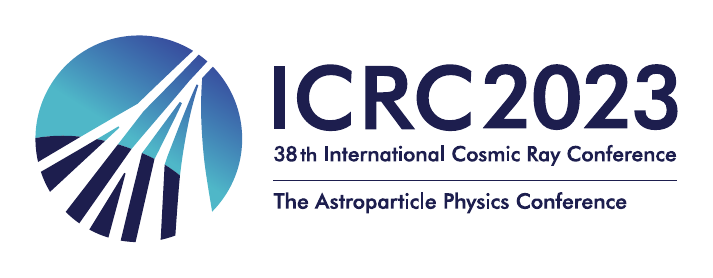}

\FullConference{%
38th International Cosmic Ray Conference (ICRC2023)\\
  26 July - 3 August, 2023\\
  Nagoya, Japan}
}

\begin{document}
\maketitle

\section{Introduction}\label{intro}
\subsection{Neutrino production mechanisms in magnetars} \label{nu emission mech}

Magnetars are neutron stars with magnetic fields of the order of $10^{13}$-$10^{15}$ G that emit strong X-ray and thermal radiation early in life due to the decay of their magnetic fields and maintain high X-ray luminosity over a long period of time ($\sim 10^4$ years) \cite{Zhang_2003}. Anomalous X-ray Pulsars (AXPs) and Soft Gamma-ray Repeaters (SGRs) are two types of magnetars with similar properties such as pulse periods, spin-down rates and quiescent X-ray luminosity. However, AXPs tend to be less active than SGRs.

Young magnetars with oppositely-oriented magnetic fields and spin moments may emit high-energy (HE) neutrinos from their polar caps as they accelerate cosmic rays (mainly protons) to high energies\cite{Zhang_2003}. Neutrino emission due to the acceleration of high-energy protons usually happens through pion decay. Pions are produced when protons interact with photons (photomeson interaction) or matter in the environment of the astrophysical accelerators. For the case of most pulsars, which are highly-magnetized, fast-spinning neutron stars, the immediate environment such as the magnetosphere lacks a target column density large enough for pion production. Therefore, the neutrino emission process in pulsars is usually expected to happen in the pulsar wind nebula \cite{Beall_2002}. However, in the inner magnetosphere of pulsars with surface magnetic fields of $\sim 10^{15}$ G, i.e. magnetars, conditions for neutrino production via photomeson interaction are realized. 

In principle, there are two main sources of energy that power a magnetar: the spin-down power which is a measure of loss of rotational energy of the magnetar, and the power resulting from decaying magnetic fields (magnetic power). The spin-down power accelerates protons and the magnetic power provides a large amount of near-surface photons. Assuming both of these energy sources power the magnetar, and that the magnetar is young enough, then the criterion for photomeson interactions are satisfied \cite{Zhang_2003}.

The dominant photomeson interaction resulting in neutrino emission in magnetars then is through the $\Delta$-resonance \cite{Zhang_2003}:

\begin{equation}
    p\gamma \rightarrow \Delta \rightarrow n\pi^{+} \rightarrow n\nu_{\mu}\mu^{+} \rightarrow n\nu_{\mu}e^{+}\nu_{e}\bar{\nu}_\mu.
\end{equation}

Post-burst magnetars (SGRs after flaring) show an increase in their quiescent luminosity for a long period of time, therefore they could contribute higher neutrino fluxes \cite{Zhang_2003}. Giant flares of SGRs may produce HE neutrinos which are potentially detectable by IceCube \cite{Ioka_2005}.

\subsection{Neutrino detection and The IceCube Neutrino Observatory}\label{icecube}

The IceCube Neutrino Observatory is a cubic-kilometer neutrino detector deep inside the Antarctic ice \cite{Aartsen_2017} which was completed in 2011 and was already operational before that while partially completed. In 2013, IceCube published the first evidence of HE neutrinos of astrophysical origin \cite{1242856}. 

As the HE neutrinos interact with the Antarctic ice, they produce relativistic charged particles which emit Cherenkov light. The detector consists of 5160 digital optical modules (DOMs) which can detect the Cherenkov light. Using the signals from the DOMs, one can infer the energy, direction, and flavor of the HE neutrino. The charged-current interactions of muon neutrinos produce high-energy muons that can travel kilometers in the ice. These muon tracks have an angular resolution of $\sim 1^{\circ}$ for energies above 10 TeV. In this work we plan to use a sample of events from both the northern and southern sky using 14 years of IceCube data, from April 2008 through May 2022, which are optimized for astrophysical neutrino point source searches \cite{PhysRevLett.124.051103}.

\section{The catalog of magnetars}\label{catalog}

\begin{figure}[!h]
    \centering
    \includegraphics[width=0.65\paperwidth]{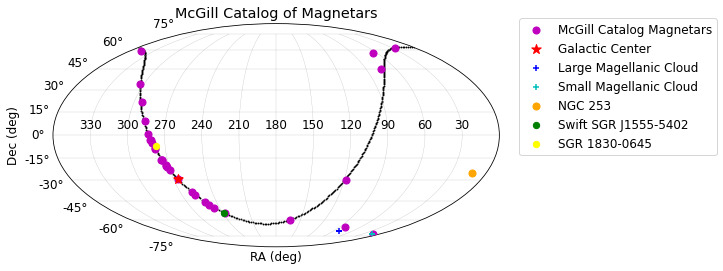}
    \caption{Position of the magnetars used in this analysis.}
    \label{fig:mag_cat}
\end{figure}

For this search, we use the McGill Online Magnetar Catalog \cite{Olausen_2014} last modified on November 17, 2020. In addition to the McGill catalog, two newly discovered magnetars, Swift J1555.2$-$5402 \cite{swiftj1555} and SGR 1830$-$0645 \cite{SGRJ1830} are included in this analysis. $\gamma$-ray burst GRB 200415A is believed to be a giant flare of a magnetar localized to a 20-square-arcmin region of the starburst galaxy NGC 253 \cite{Svinkin2021}. Since the angular resolution of IceCube is about 1$^{\circ}$, as discussed earlier, we assume the position of the magnetar to be the position of the galaxy itself.
PSR J1846$-$0258 is excluded from the stacked analysis (see section \ref{stacked search}) since it is often classified as a young, rotation-powered pulsar. Rotation-powered pulsars as the name suggests are not powered by the decay of their magnetic fields. This particular pulsar has a magnetic field strength much higher than rotation-powered pulsars and in 2006 was seen to have undergone a magnetar-like outburst \cite{doi:10.1126/science.1153465}. The positions of the magnetars included in this work are shown in Fig.\ref{fig:mag_cat}.

\section{Analysis}

\subsection{Point source search}
 We want to determine if any of the events in our data originate from magnetars or if they are all due to background. We will use statistical hypothesis testing, with the null hypothesis $\mathcal{H}_0$ being all events are due to background, and the alternate hypothesis $\mathcal{H}_1$ being the events are due to both background and an astrophysical neutrino signal. We will model the data and determine the likelihood of each hypothesis using the test statistics (TS), defined by

\begin{equation}
    \mathcal{TS} = -2\log \left[\frac{P(data|\mathcal{H}_0)}{P(data|\mathcal{H}_1)} \right]\,.
    \label{ts}
\end{equation}

Larger values of TS mean the null hypothesis is less likely.  As stated in Wilks' Theorem \cite{wilks}, in cases where the null hypothesis is true, values of the TS will follow a chi-squared ($\chi^2$) distribution with $m$ degrees of freedom depending on the number of fit parameters, in this case 2.

We use the unbinned likelihood method to search for neutrino point sources. The likelihood function for a single point source is given by: 
\begin{equation}
    \mathcal{L}(x_s,n_s) = \prod_{i = 1}^{N}  \left(\frac{n_s}{N}\mathcal{S}(x_i,x_s,E_i,\gamma) +\left(1 -\frac{n_s}{N}\right)\mathcal{B}(x_i, E_i)\right),
    \label{likelihood}
\end{equation}

\noindent where $x_s$ is the position of the source, $n_s$ is the number of the signal events, $x_i$ and $E_i$ are the reconstructed position and energy of the $i$th neutrino candidate event (hereafter event), $N$ is the total number of events, and $\gamma$ is the spectral index.

The background probability density function (PDF) $\mathcal{B}$ is a function of the reconstructed energy and the declination of the events. The background PDF does not depend on the right ascension (RA) since the effective area of the detector, averaged over time, is constant with respect to RA. This does not take into account the diffuse emission from the Galactic plane \cite{doi:10.1126/science.adc9818}. The background PDF is given by:
\begin{equation}
\mathcal{B}_i(\vec{x}_i,E_i) = \underbrace{B_i(\vec{x}_i)}_{\text{Spatial}} \times \underbrace{\mathcal{E_B}_i(E_i)}_{\text{Energy}}.
\label{bg pdf}
\end{equation}

The signal PDF  $\mathcal{S}$ is also assumed to only have spatial and energy components for the time-integrated search and to be Gaussian in form which describes the point-spread function of the IceCube detector. For a given source:

\begin{equation}
\mathcal{S}_i(\vec{x}_i,\vec{x}_{S},\sigma_i,E_i,\gamma) = \underbrace{S_i(\vec{x}_i,\vec{x}_{S},\sigma_i)}_{\text{Spatial}} \times \underbrace{\mathcal{E_S}_i(E_i,\gamma)}_{\text{Energy}}  .  
\label{sig pdf}
\end{equation}
\noindent $S_i$ is described below as:

\begin{equation}
S_i(\vec{x}_{i},\vec{x}\textsubscript{s},\sigma_i) = \frac{1}{2 \pi \sigma_i^2} \cdot \exp \left(- \frac{| \vec{x}_i - \vec{x}\textsubscript{s} |^2}{2\sigma_i^2}\right),
\label{spatial sig}
\end{equation}

where $\sigma_i$ is the reconstructed directional uncertainty of each event. $\mathcal{E_S}_i(E_i,\gamma)$ are reconstructed energies.

\subsection{Time-integrated stacking analysis}\label{stacked search}

In a stacking analysis contributions from all the sources are added up (stacked). Therefore if the neutrino flux from a single magnetar is not detectable, it still contributes to the total neutrinos emitted from magnetars. This allows us to evaluate neutrino emission from magnetars as a class of objects. The likelihood function for the stacked analysis takes the same form as Eq. \ref{likelihood}, but with a modified signal PDF given by:

\begin{equation}
    \mathcal{S}_{i,Stacked} =
      \frac{\sum_{j=1}^M w_j R_j(\delta_j,\gamma) \cdot
        \mathcal{S}_i,j(\vec{x}_i,\vec{x}_{S_j},\sigma_i,E_i, \gamma)}{\sum_{j=1}^M w_j R_j(\delta_j,\gamma)}. 
\end{equation}

where the index $j$ denotes the source from our catalog, and $w_j$ is the theoretical weight depending on physical properties of the source and $R(\delta_j,\gamma)$ is the detector weight describing the sensitivity of the IceCube detector for a source at a declination $\delta $ with a spectral index $\gamma$.

Two choices of weights, based on different theoretical models of neutrino emission, are tested:
\begin{itemize}
    
    \item \textbf{Energy flux}: Neutrino flux and the unabsorbed X-ray energy flux have a direct correlation.

    \item \textbf{Spindown $\frac{\dot P}{P^3}$}: Young magnetars are more likely to emit high energy neutrinos. As magnetars age, their periods increase.
\end{itemize}

\section{Sensitivity and discovery potentials}

\begin{figure}[h!]
    \centering
    \begin{subfigure}[t]{0.49\textwidth}
        \centering
        \includegraphics[width=0.95\textwidth]{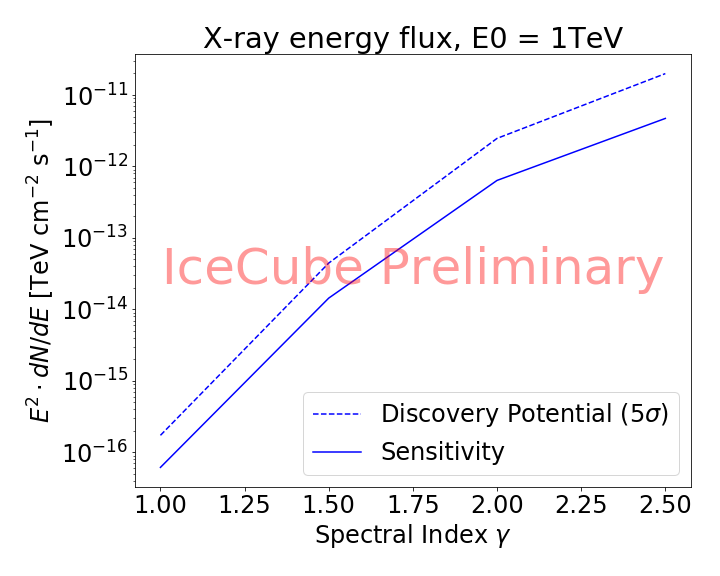}

    \end{subfigure}
    ~
    \begin{subfigure}[t]{0.49\textwidth}
        \centering
        \includegraphics[width=0.95\textwidth]{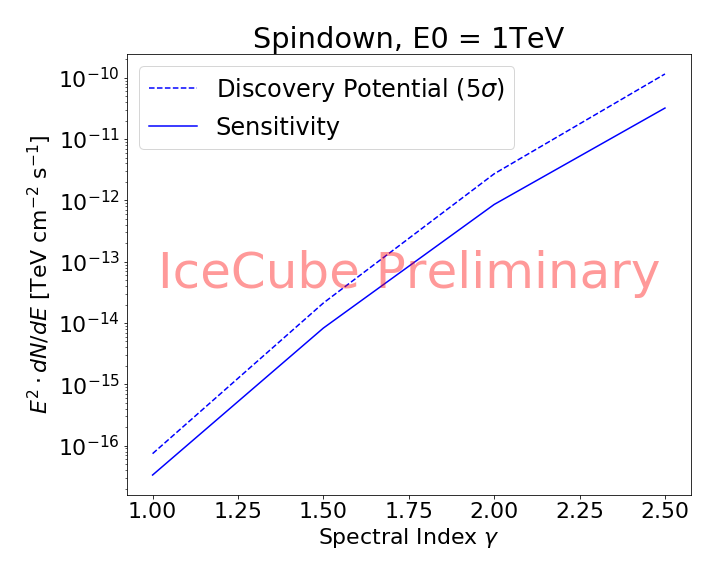}
    \end{subfigure}
    
    \caption{Sensitivity and $5\sigma$ discovery potential curves with (left) x-ray energy flux, (right) spindown as weights with $E_0=1$ TeV. The flux is the total flux resulting from stacking.}
    \label{fig:flux sens}
\end{figure}

The significance of an analysis result is determined by estimating how often such a result of similar or greater strength would occur by chance in an analysis of background-only data. This estimate is obtained by performing the analysis repeatedly with the same data but having the right ascension coordinate randomized (called "scrambled trials"). The distribution of the background TS values obtained by performing many scrambled trials (representing the null hypothesis) should follow a $\chi^2$ distribution. The PDF is obtained by performing pseudo-experiments where Monte Carlo simulated neutrino events are injected at the position of the magnetars. The injected neutrino signal has a power law spectrum with a spectral index $\gamma_{inj}$:

\begin{equation}
\frac{dN}{dE}=N\left({\frac {E}{E_{0}}}\right)^{-\gamma_{inj} },
\end{equation}

where $\frac{dN}{dE}$ is the differential neutrino flux as a function of energy, normalized at $E_0$ = 1 TeV. In this work, we have injected signals with spectral indices of $1.0,\ 1.5,\ 2.0,\ 2.5$. Figure \ref{fig:flux sens} shows the sensitivity and discovery potential for the analyses described here. We define sensitivity as number of signal events needed to have a TS greater than the background median in $90\%$ of the trials performed. The discovery potential is the number of injected signal events needed to reach $5\sigma$ in $50\%$ of the performed trials. 

\newpage
\section{Discussion}
\begin{figure}[h!]   
\centering
    \begin{subfigure}[t]{0.65\textwidth}
        \centering
        \includegraphics[width=0.95\textwidth]{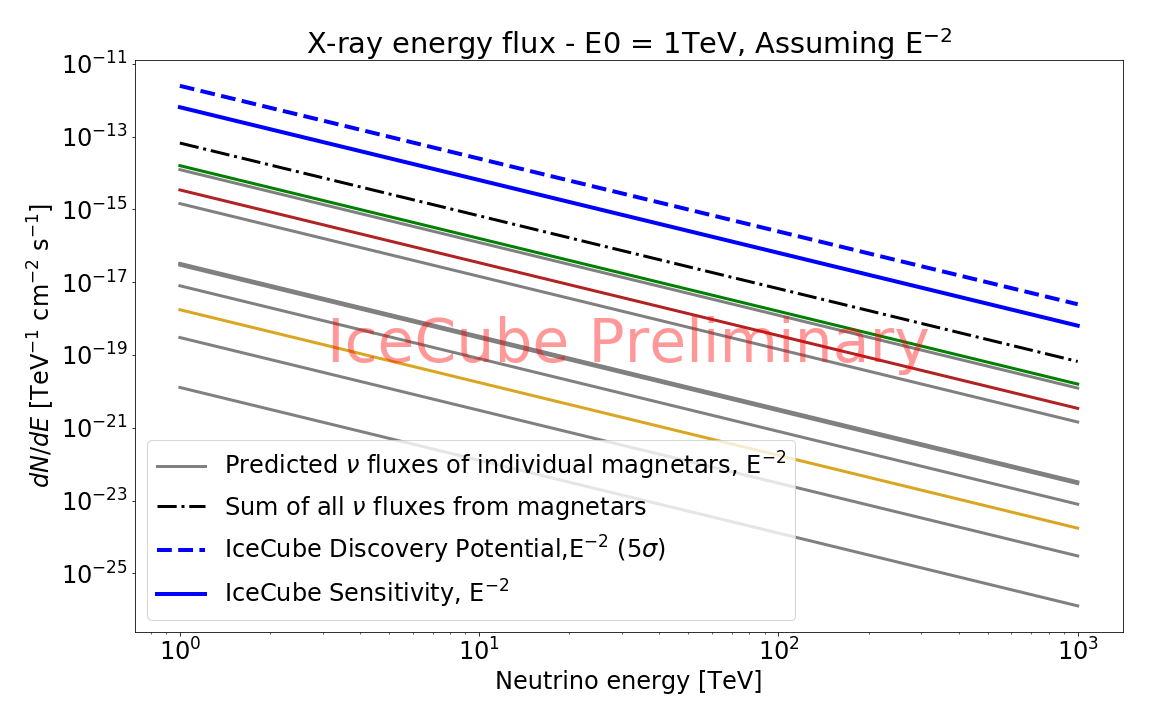}
    \end{subfigure}
    ~
    \begin{subfigure}[t]{0.65\textwidth}
        \centering
        \includegraphics[width=0.95\textwidth]{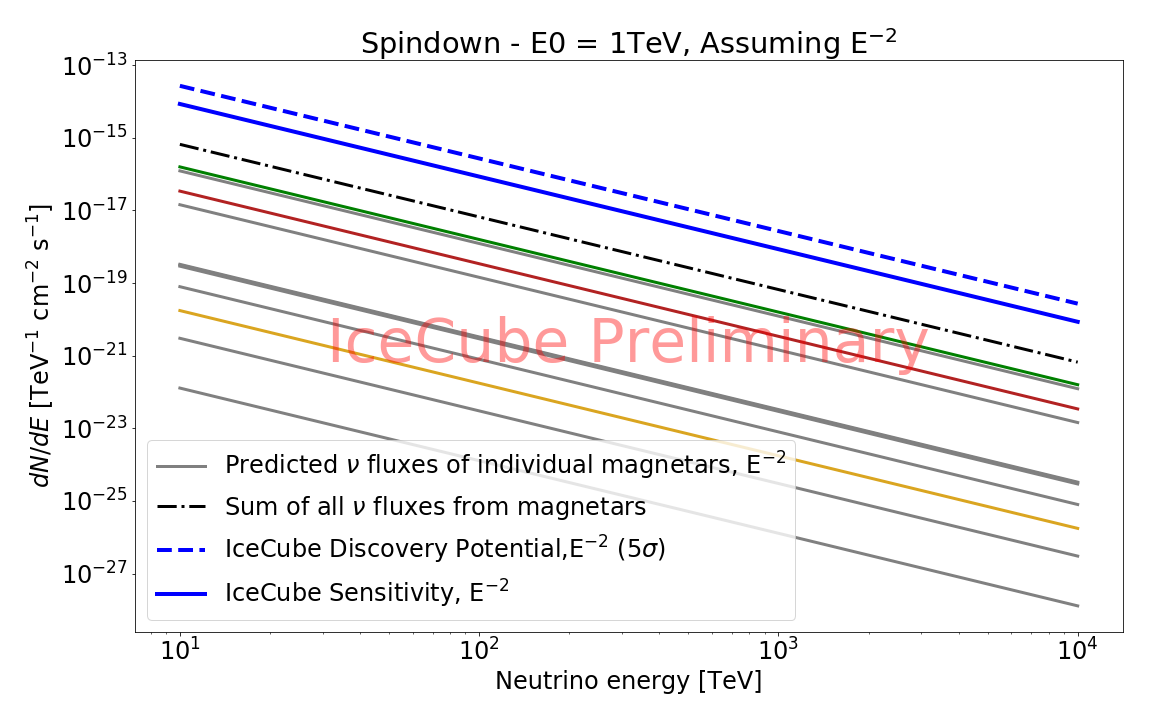}
    \end{subfigure}

    \caption{ Differential neutrino flux of magnetars derived from Zhang et. al. compared to the sensitivity and discovery potential of IceCube, using (top) x-ray energy flux weighting and (bottom) spindown weighting, gamma=2.0. Most individual predicted neutrino fluxes are shown with grey solid lines; colored solid lines indicate the predicted neutrino fluxes of Swift J1834.9-0846 (dark green), SGR 1935+2154 (dark red), and Swift J1818.0-1607 (dark yellow) according to \cite{Zhang_2003}. }
    \label{fig:sens_disc_zhang}
\end{figure}

Zhang \textit{et al.} first proposed a model that describes high-energy neutrino emission from magnetars in 2003 \cite{Zhang_2003}. As described in section \ref{nu emission mech}, young magnetars with oppositely-oriented magnetic fields may emit high-energy neutrinos from their polar caps. The neutrino number flux arriving at Earth is predicted by Equation 15 in \cite{Zhang_2003}. Taking the derivative $\frac{d\phi_{\nu}}{d\epsilon_{\nu}}$ gives us the differential neutrino flux from the magnetars. Using the data in the McGill magnetar catalog \cite{Olausen_2014}, we have plotted this model and compared it to our sensitivity and discovery potentials in Fig. \ref{fig:sens_disc_zhang}. The sum of the neutrino fluxes from the magnetars in the catalog is below the IceCube sensitivity derived in this work for both weighting schemes.

\section{Future plans}
Thus far we have used Monte Carlo (MC) simulated events to investigate the hypothesis that magnetars are potential HE neutrino emitters. We have produced sensitivity and $5\sigma$ discovery potential curves for both stacking and point source analyses. Upon completing the internal review procedure, we will perform the analysis on the real data (unblind) and will report the result in a future publication. 
\\

The transient outburst of magnetars is a key property of their emission and has played a major role in their discovery \cite{Rea_2010}. The exact mechanism responsible for the outbursts however, is not clear \cite{10.1093/mnras/stx2679}. The outbursts are defined as periods where the persistent emission increases by a factor of 10-1000 and lasts from a few weeks to a couple of years. It is worth investigating whether the outburst of magnetars results in neutrino emission. We plan to conduct a transient search for HE neutrinos from the outbursts of magnetars with IceCube as the next step in our pursuit to better understand these objects.

\bibliographystyle{ICRC}
\bibliography{references}

\newpage

\input{authorlist_IceCube.tex}

\end{document}

%% file: authorlist_IceCube.tex
\section*{Full Author List: IceCube Collaboration}

\scriptsize
\noindent
R. Abbasi$^{17}$,
M. Ackermann$^{63}$,
J. Adams$^{18}$,
S. K. Agarwalla$^{40,\: 64}$,
J. A. Aguilar$^{12}$,
M. Ahlers$^{22}$,
J.M. Alameddine$^{23}$,
N. M. Amin$^{44}$,
K. Andeen$^{42}$,
G. Anton$^{26}$
% ...
\\
\\
% ...
$^{12}$ Universit{\'e} Libre de Bruxelles, Science Faculty CP230, B-1050 Brussels, Belgium \\
% ...
$^{17}$ Department of Physics, Loyola University Chicago, Chicago, IL 60660, USA \\
$^{18}$ Dept. of Physics and Astronomy, University of Canterbury, Private Bag 4800, Christchurch, New Zealand \\
% ...
$^{22}$ Niels Bohr Institute, University of Copenhagen, DK-2100 Copenhagen, Denmark \\
$^{23}$ Dept. of Physics, TU Dortmund University, D-44221 Dortmund, Germany \\
% ...
$^{26}$ Erlangen Centre for Astroparticle Physics, Friedrich-Alexander-Universit{\"a}t Erlangen-N{\"u}rnberg, D-91058 Erlangen, Germany \\
% ...
$^{40}$ Dept. of Physics and Wisconsin IceCube Particle Astrophysics Center, University of Wisconsin{\textendash}Madison, Madison, WI 53706, USA \\
% ...
$^{42}$ Department of Physics, Marquette University, Milwaukee, WI, 53201, USA \\
% ...
$^{44}$ Bartol Research Institute and Dept. of Physics and Astronomy, University of Delaware, Newark, DE 19716, USA \\
% ...
$^{63}$ Deutsches Elektronen-Synchrotron DESY, Platanenallee 6, 15738 Zeuthen, Germany  \\
$^{64}$ Institute of Physics, Sachivalaya Marg, Sainik School Post, Bhubaneswar 751005, India \\
% ...

\subsection*{Acknowledgements}

\noindent
The authors gratefully acknowledge the support from the following agencies and institutions:
USA {\textendash} U.S. National Science Foundation-Office of Polar Programs,
U.S. National Science Foundation-Physics Division,
U.S. National Science Foundation-EPSCoR,
Wisconsin Alumni Research Foundation,
Center for High Throughput Computing (CHTC) at the University of Wisconsin{\textendash}Madison,
Open Science Grid (OSG),
Advanced Cyberinfrastructure Coordination Ecosystem: Services {\&} Support (ACCESS),
Frontera computing project at the Texas Advanced Computing Center,
U.S. Department of Energy-National Energy Research Scientific Computing Center,
Particle astrophysics research computing center at the University of Maryland,
Institute for Cyber-Enabled Research at Michigan State University,
and Astroparticle physics computational facility at Marquette University;
Belgium {\textendash} Funds for Scientific Research (FRS-FNRS and FWO),
FWO Odysseus and Big Science programmes,
and Belgian Federal Science Policy Office (Belspo);
Germany {\textendash} Bundesministerium f{\"u}r Bildung und Forschung (BMBF),
Deutsche Forschungsgemeinschaft (DFG),
Helmholtz Alliance for Astroparticle Physics (HAP),
Initiative and Networking Fund of the Helmholtz Association,
Deutsches Elektronen Synchrotron (DESY),
and High Performance Computing cluster of the RWTH Aachen;
Sweden {\textendash} Swedish Research Council,
Swedish Polar Research Secretariat,
Swedish National Infrastructure for Computing (SNIC),
and Knut and Alice Wallenberg Foundation;
European Union {\textendash} EGI Advanced Computing for research;
Australia {\textendash} Australian Research Council;
Canada {\textendash} Natural Sciences and Engineering Research Council of Canada,
Calcul Qu{\'e}bec, Compute Ontario, Canada Foundation for Innovation, WestGrid, and Compute Canada;
Denmark {\textendash} Villum Fonden, Carlsberg Foundation, and European Commission;
New Zealand {\textendash} Marsden Fund;
Japan {\textendash} Japan Society for Promotion of Science (JSPS)
and Institute for Global Prominent Research (IGPR) of Chiba University;
Korea {\textendash} National Research Foundation of Korea (NRF);
Switzerland {\textendash} Swiss National Science Foundation (SNSF);
United Kingdom {\textendash} Department of Physics, University of Oxford.